\documentclass[conference]{IEEEtran}
\usepackage{cite}
\usepackage{graphicx}
\usepackage{multirow}
\usepackage{url}
\usepackage[caption=false,font=footnotesize]{subfig}
\usepackage{booktabs}
\usepackage{array}
\newcolumntype{C}[1]{>{\centering\arraybackslash}p{#1}} % centered p-column

\usepackage{amsthm}
\usepackage[cmex10]{amsmath}
\usepackage{comment}
\usepackage{amsfonts}
\usepackage{amssymb}
\graphicspath{{figures/}}
\usepackage{bm}
\usepackage{epstopdf}
\usepackage{float}
\usepackage{xcolor}

\usepackage[T1]{fontenc}
\newcommand*{\affmark}[1][*]{\textsuperscript{\dag}}
\usepackage{algorithm}
\usepackage[noend]{algpseudocode}
\makeatletter
\def\BState{\State\hskip-\ALG@thistlm}
\makeatother
\usepackage{acronym}
\acrodef{ISAC}[ISAC]{Integrated Sensing and Communication}
\definecolor{revisionred}{RGB}{180,0,0}

\newcommand{\panelcaption}[1]{\vspace{0.2em}\centerline{\footnotesize \textcolor{black}{#1}}}

\ifCLASSINFOpdf
  \else  
\fi

\usepackage[normalem]{ulem}

\begin{document}
\title{Multi-TRP Assisted UAV Detection in 3GPP 5G-Advanced ISAC Network}

\author{\IEEEauthorblockN{Neeraj Varshney$^{1,2}$, Steve Blandino$^{1,2}$, Jian Wang$^{1}$, Anuraag Bodi$^{1,2}$, Camillo Gentile$^{1}$, and Nada Golmie$^{1}$}
\IEEEauthorblockA{$^{1}$National Institute of Standards and Technology
Gaithersburg, Maryland USA\\
$^{2}$Prometheus Computing LLC, Bethesda, MD.}}


%Telephone: (800) 555--1212, Fax: (888) 555--1212}


\maketitle

\begin{abstract}
Integrated Sensing and Communication (ISAC) is currently being standardized within the {Third Generation Partnership Project (3GPP)} New Radio (NR) to enable cellular infrastructure to perform sensing using existing communication waveforms. While standardization is progressing, practical deployment may be limited by scenario-dependent observability constraints. For example, in Urban Macro Aerial Vehicle (UMa-AV) scenarios, sensing with a single transmission-reception point (TRP) can be affected by restricted angular coverage, partial blockage, and limited field of view, which may degrade detection reliability in three-dimensional Unmanned Aerial Vehicle (UAV) environments. For this reason, multi-TRP solutions have been suggested to improve spatial diversity and sensing robustness. In this paper, we present a system-level investigation of multi-TRP assisted monostatic sensing for UAV detection under standardized 3GPP UMa-AV channel assumptions and Release 19 evaluation parameters. We propose a spatial diversity fusion framework and evaluate the achievable performance of a 3GPP network by combining the measurements obtained independently at different TRP. Extensive evaluations demonstrate that multi-TRP assistance improves target observability, reduces spurious detections, and tightens localization error distributions at the cost of additional sensing overhead due to the need for multiple TRPs to periodically allocate radio resources for sensing measurements. In the evaluated scenario, results show that a voting threshold of two assisting TRPs achieves an optimal trade-off between miss detection probability and false alarm suppression, meeting 3GPP performance objectives. Furthermore, we quantify the sensing overhead and show that proper system design, tuned to the application requirements, can substantially reduce its impact: for example, extending the sensing refresh interval beyond the 128 ms coherent processing interval to 1 s reduces the effective overhead from 29~\% to approximately 3.7~\%, enabling more scalable network deployment.
\end{abstract}


% \fi
\IEEEpeerreviewmaketitle

\section{Introduction}
Integrated Sensing and Communication (ISAC) has emerged as a key technology direction for 5G-advanced/6G networks, enabling wireless infrastructure to perform environmental sensing while maintaining conventional communication services  \cite{qaisar2026role}. Recognizing this potential, {the Third Generation Partnership Project (3GPP)} has initiated study items on ISAC for New Radio (NR) in Release 19 and beyond, focusing on deployment scenarios such as the Urban Macro Aerial Vehicle (UMa-AV) environment where network infrastructure performs sensing using standardized reference signals \cite{3GPP38765}. In this configuration, macro base stations ({next generation NodeB (gNB)} or {transmission-reception point (TRP)}) detect and track Unmanned Aerial Vehicles (UAVs) operating at varying altitudes, utilizing channel characteristics and 3D propagation models standardized in {Technical Report (TR)} 38.901 \cite{3GPP38901}. 

UAV detection in UMa-AV environments presents distinct challenges due to three-dimensional (3D) motion, especially flying above the gNB at high altitudes \cite{blandino2025detecting}. In such complex 3D environments with dynamic transitions between Line-of-Sight (LOS) and Non-LOS (NLOS) channels a single serving TRP may fail to detect the presence of the UAV or suffer from fundamental geometric observability limitations \cite{dickerson2025impact,tang2025uav}. To address these issues, multi-TRP assistance provides a natural solution by combining independent observations from spatially separated TRPs to improve detection reliability and localization consistency \cite{semenyuk2025advances}.

Within the 3GPP framework, the 5G NR Positioning Reference Signal (PRS) has demonstrated significant potential for high-resolution delay and Doppler estimation in monostatic configurations \cite{bednarz2025remote,sagduyu2025multi}. More recently, research has shifted toward distributed sensing architectures where spatially separated nodes cooperatively utilize the same PRS waveform for enhanced target detection and localization \cite{sagduyu2025multi, Demir2024IScout, Khosroshahi2024Multistatic}. However, comprehensive system-level analysis under standardized 3GPP UMa-AV assumptions remains scarce. In particular, practical implementation challenges such as multi-TRP target association, spatial clustering, and resource-efficient voting strategies within the context of the TR 38.901 ISAC channel have not been fully addressed.  

This paper therefore  addresses these challenges by presenting a system-level investigation of multi-TRP assisted sensing. We present a practical framework combining Cartesian clustering, voting-based target confirmation, and geometry-consistent least-squares velocity reconstruction. Note that radial velocity measurements from spatially separated TRPs vary, as they are projections of the true 3D target motion onto different observation vectors. Consequently, we employ a geometry-consistent least-squares reconstruction to synthesize these diverse observations into a 3D velocity vector. This allows the system to accurately estimate the target's radial velocity relative to the central fusion TRP, ensuring spatial consistency across the 3GPP network.  Extensive evaluations demonstrate that multi-TRP assistance improves target observability, reduces spurious detections, and tightens localization error distributions at the cost of additional sensing overhead due to the need for multiple TRPs to periodically allocate radio resources for sensing measurements. Therefore, we also study the impact on overhead.

% Our key contributions are as follows:
% \begin{enumerate}
%     \item \textit{Fusion Framework:} We presents a practical framework combining Cartesian clustering, voting-based target confirmation, and geometry-consistent least-squares velocity reconstruction.
%     \item \textit{Optimal Voting Threshold:} We identify that a voting threshold of two assisting TRPs provides the optimal trade-off between suppressing false alarms and maintaining a low miss detection probability ($<5\%$), ensuring robust performance even at low transmit power.
%     \item \textit{Velocity Estimation Optimization:} Unlike horizontal and vertical position errors which consistently decrease as the number of assisting TRPs increases to four, we demonstrate that radial velocity estimation is optimized when using the three strongest TRPs, as the fourth unreliable TRP can introduce geometric noise into the least-squares reconstruction.
%     \item \textit{Overhead Reduction:} Based on the 3GPP agreement, we also quantify that extending refresh intervals beyond the coherent processing interval (CPI) reduces sensing overhead to approximately 3.7\%.
% \end{enumerate}

The remainder of this paper is organized as follows: Section~\ref{Sec:SystemModel} introduces the 5G NR monostatic radar system model including 3GPP Release-19 ISAC channel model. Section~\ref{Sec:MultiTRP} describes the fusion framework. Section~\ref{Sec:EvaluationResults} presents the evaluation results and performance insights. Finally, Section~\ref{Sec:Conclusion} concludes the paper.

\begin{figure}[t]
\centering
\includegraphics[width=0.45\linewidth]{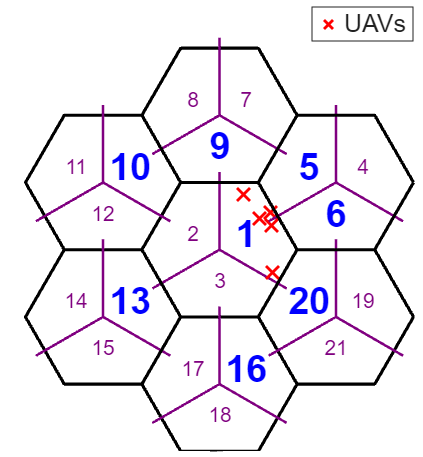}
\vspace{-0.2cm}
\caption{Hexagonal multi-site deployment for the 3GPP NR ISAC UMa-AV scenario, illustrating the TRP sector numbering and spatial distribution of UAV targets, shown as red $\times$'s.}
\label{fig:deployment}
\vspace{-0.5cm}
\end{figure}

\vspace{-0.1cm}
\section{System Model}
\vspace{-0.1cm}
\label{Sec:SystemModel}
In this work, we consider a monostatic ISAC system consisting of $N_{\mathrm{TRP}}$ spatially distributed TRPs. The deployment follows the 3GPP NR ISAC UMa-AV scenario with $N_{\mathrm{site}}=7$ macro cell sites arranged in a hexagonal layout, as illustrated in Fig.~\ref{fig:deployment}. Each site is equipped with three sectorized antennas providing approximately $120^\circ$ coverage per sector, resulting in a total of $N_{\mathrm{TRP}}=21$ sectors. 
%Each sector acts as an independent monostatic sensing node. 
The TRPs are indexed by $i \in \{1,\cdots,N_{\mathrm{TRP}}\}$ according to the sector numbering shown in Fig.~\ref{fig:deployment}. Each TRP$i$ consists of co-located transmit and receive antenna arrays with $N$ elements arranged in a uniform rectangular array (URA) and operates in monostatic sensing mode using a Cyclic Prefix-Orthogonal Frequency Division Multiplexing (CP-OFDM) waveform. The system employs downlink PRS as the sensing signal, enabling delay and Doppler estimation for target detection and localization. In this study, PRS resources are exclusively utilized for sensing evaluation, and communication traffic is not considered. 
\subsection{Transmit Signal Model}
Let $\Delta f$ denote the subcarrier spacing and $N_{\mathrm{sc}}$ the number of active subcarriers. The PRS symbols are mapped on the OFDM grid such that only every $K$th subcarrier carries a reference symbol. The transmitted signal from TRP$i$ on subcarrier $k$ and OFDM symbol $\ell$ can be written as
\begin{equation}
\mathbf{x}_{i,k,\ell} = \sqrt{P_{\mathrm{TX}}}\,\mathbf{f}_i\,s_{k,\ell} \in \mathbb{C}^{N\times1},
\end{equation}
where $P_{\mathrm{TX}}$ denotes the transmit power per resource element, $\mathbf{f}_i \in \mathbb{C}^{N\times1}$ represents the transmit precoder at TRP$i$, and $s_{k,\ell}$ is the known PRS symbol. The precoder is designed to provide quasi-omnidirectional coverage within the sensing sector. 

\subsection{ISAC Channel Model}

The propagation channel between TRP$i$ and the environment consists of reflections from sensing targets and background clutter. The baseband-equivalent time-varying {multiple-input multiple-output (MIMO)} channel impulse response can therefore be expressed as
\begin{equation}
\mathbf{H}_i(\tau,t) = \mathbf{H}^{(\mathrm{ST})}_i(\tau,t) + \mathbf{H}^{(\mathrm{BG})}_i(\tau,t),
\end{equation}
where $\mathbf{H}^{(\mathrm{ST})}_i(\tau,t)$ represents the sensing target channel and $\mathbf{H}^{(\mathrm{BG})}_i(\tau,t)$ denotes the background clutter channel \cite{3GPP38901}. 

Each sensing target is modeled as one scattering point characterized by radar cross section (RCS). For $Q$ targets, the monostatic target channel observed at TRP$i$ can be written as
\begin{align}
\mathbf{H}^{(\mathrm{ST})}_i(\tau,t)
=
&\sum_{q=1}^{Q}
\sum_{p=1}^{P_q}
\alpha_{i,q,p}(t)
\mathbf{a}_{\mathrm{rx},i}(\Omega^{\mathrm{AOA}}_{i,q,p}) \nonumber \\
&\times\mathbf{a}_{\mathrm{tx},i}^{H}(\Omega^{\mathrm{AOD}}_{i,q,p})
\delta(\tau-\tau_{i,q,p}),
\end{align}
where $H$ is the Hermitian (complex conjugate transpose) operator, $P_q$ denotes the number of multipath components associated with target $q$, $\tau_{i,q,p}$ represents the round-trip propagation delay between TRP$i$ and the corresponding scatterer. The quantities  $\Omega^{\mathrm{AOA}}_{i,q,p}$ and $\Omega^{\mathrm{AOD}}_{i,q,p}$ represent the {angle of arrival (AoA)} and {angle of departure (AoD)}, respectively, and  $\mathbf{a}_{\mathrm{rx},i}(\cdot)$ and $\mathbf{a}_{\mathrm{tx},i}(\cdot)$ denote receive and transmit array response vectors, whereas $\alpha_{i,q,p}(t)$ represents the complex path gain incorporating path loss, shadow fading, radar cross section, and Doppler phase due to target motion.

The background channel models clutter reflections and is represented as
\begin{equation}
\textbf{H}^{(\mathrm{BG})}_i(\tau)
=
\sum_{r=1}^{N_{\mathrm{RP}}}
10^{-(PL_{i,r}+SF_{i,r})/20}
\textbf{H}^{(\mathrm{RP})}_{i,r}(\tau), \label{bgChannel}
\end{equation}
where $N_{\mathrm{RP}}$ denotes the number of virtual reference points representing clutter scatterers, while $PL_{i,r}$ and $SF_{i,r}$ represent path loss and shadow fading associated with the $r$th reference point with respect to TRP$i$. $\textbf{H}^{(\mathrm{RP})}_{i,r}(\tau)$ is the time-invariant channel model for the $r$th reference point for TRP$i$. Note that for a stationary TRP and static virtual reference points, the background has no geometric time evolution and is therefore constant over a coherent processing interval (CPI); accordingly, we treat $\textbf{H}^{(BG)}(\tau,t)=\textbf{H}^{(BG)}(\tau)$ and omit the explicit time dependence in Eq. \eqref{bgChannel}. 

\subsection{{Received Signal Model}}

After cyclic-prefix removal and OFDM demodulation, the received signal at the {radio frequency (RF)} chains of TRP$i$ can be written as
\begin{equation}
\mathbf{y}_{i,k,\ell}
=  \textbf{H}_{i,k,\ell}\mathbf{x}_{i,k,\ell}
+ \mathbf{n}_{i,k,\ell},
\end{equation}
where $\mathbf{n}_{i,k,\ell}\sim\mathcal{CN}(\bm 0,\sigma_n^2\bm I)$ represents complex-valued zero-mean white thermal noise with variance $\sigma_n^2$.

%, and $\boldsymbol{\xi}_{i,k,\ell}\sim\mathcal{CN}(\bm 0,\sigma_{si}^2\bm I)$ denotes residual self-interference (SI) leakage after cancellation.

\subsection{{ISAC Signal Processing}}
For sensing processing, the received PRS symbols are collected over a CPI. Each PRS occasion contains $L_{\mathrm{PRS}}$ PRS OFDM symbols and repeats every $T_{\mathrm{PRS}}$ seconds. A CPI consists of $N_{\mathrm{CPI}}$ consecutive PRS occasions, yielding a total CPI duration of $T_{\mathrm{CPI}} = N_{\mathrm{CPI}} T_{\mathrm{PRS}}$. The sensing receiver at TRP$i$ processes the received PRS symbols to estimate target parameters. First, channel estimation is performed using a least-squares estimator
\begin{equation}
\hat{\mathbf{g}}_{i,k,\ell}
=
\frac{1}{\sqrt{P_{\mathrm{TX}}}}
\frac{\mathbf{y}_{i,k,\ell}s_{k,\ell}^{*}}{|s_{k,\ell}|^{2}}.
\end{equation}
Next, range processing is performed independently on each RF chain. This is carried out by applying an $N_R$-point inverse Fast Fourier Transform (IFFT) across the frequency domain to obtain the delay profile
\begin{equation}
{r_i^{}[n,m]}
=
\frac{1}{\sqrt{N_R}}
\sum_{k}
w_R[k]{\hat{g}_{i,k,m}^{}} e^{j2\pi kn/N_R}.
\end{equation}
where $w_R[k]$ is a window function, and $N_R$ denotes the number of range bins.

Subsequently, Doppler processing is performed across slow-time samples using an Fast Fourier Transform (FFT) to generate a range–Doppler map. Static clutter is suppressed via slow-time mean subtraction. Target detection is then performed using constant false alarm rate (CFAR) detection, followed by angle estimation and 3D localization of detected targets\cite{blandino2025detecting}.  For each successful detection, TRP$i$ reports a set of measurement vectors to a central fusion node, given by $\widehat{\mathbf{m}}_i = [x_i, y_i, z_i, v_{r,i}, \gamma_i],$ where $(x_i, y_i, z_i)$ denote the estimated 3D position of the detected target, $v_{r,i}$ is the measured radial velocity, and $\gamma_i$ is the corresponding signal-to-noise ratio (SNR). The SNR is computed as the ratio of the peak magnitude in the range–Doppler map to the noise level estimated by the CFAR detector. 

%In general, each TRP may report multiple detections per CPI, and these detections may include both true targets and spurious measurements due to noise, clutter, or multipath effects. These per-TRP measurements serve as the input to the multi-TRP fusion framework described in the next section.

\section{Multiple TRP (Multi-TRP) Assisted Sensing}
\label{Sec:MultiTRP}
Based on the signal processing described in Section~\ref{Sec:SystemModel}, each TRP provides a set of local detections represented by measurement vectors $\widehat{\mathbf{m}}_i$. These measurements are obtained independently at each TRP and are subject to detection errors, missed detections, and false alarms due to noise, clutter, and limited sensing coverage. In practical deployments, sensing with a single serving TRP is often insufficient due to restricted angular coverage, blockage, and reduced target observability, particularly in UAV scenarios with significant elevation variation (e.g., zenith blind spots). As a result, multiple TRPs may observe the same target from different spatial perspectives, leading to multiple, potentially inconsistent detections of the same object.

Given these per-TRP measurements, multi-TRP assisted sensing aims to jointly process the information from multiple TRPs to improve detection reliability and estimation accuracy. This gives rise to the following key challenges:

\begin{enumerate}
    \item \textit{TRP selection:} Given measurements from $N_{\mathrm{TRP}}$ TRPs, determine a subset of TRPs that provide reliable and geometrically diverse observations of the target.
    
    \item \textit{Target association:} Given detections $\widehat{\mathbf{m}}_i$ from multiple selected TRPs, determine which measurements correspond to the same physical target.
   
    \item \textit{Detection and parameter estimation:} Given associated detections across TRPs, determine whether the associated measurements correspond to a valid target. For validated targets, estimate the target parameters, including position and velocity, in a consistent and robust manner.
\end{enumerate}

% \begin{enumerate}
%     \item \textit{TRP selection:} Given measurements from $N_{\mathrm{TRP}}$ TRPs, determine a subset of TRPs that provide reliable and geometrically diverse observations of the target.
    
%     \item \textit{Target association:} Given detections $\hat{\mathbf{z}}_i$ from multiple selected TRPs, determine which measurements correspond to the same physical target through spatial clustering.
    
%     \item \textit{Target detection (multi-TRP validation):} Given candidate clusters formed in the association step, determine whether a cluster corresponds to a valid target using a multi-TRP voting criterion based on the number of supporting TRPs.
    
%     \item \textit{Parameter estimation:} Given validated multi-TRP detections, estimate the target parameters, including position and velocity, in a consistent and robust manner.
% \end{enumerate}

To address these challenges, we propose a multi-TRP fusion framework consisting of TRP selection, spatial clustering-based association, voting-based detection, and parameter estimation, as described in the following subsections.

\subsection{TRP Selection}

Based on the per-TRP measurements $\widehat{\mathbf{m}}_i$ obtained in Section ~\ref{Sec:SystemModel}, not all TRPs contribute equally to sensing performance. The effectiveness of multi-TRP assistance depends strongly on the geometric diversity and reliability of the available observations. In the considered 7-site sectorized deployment shown in Fig.~\ref{fig:deployment}, each TRP provides detections only within its sensing field-of-view (FOV). Consequently, a UAV may be detected by only a subset of TRPs, while others fail to observe the target due to unfavorable beam orientation, blockage, or limited vertical coverage (e.g., zenith blind spots). Therefore, the first step in multi-TRP assisted sensing is to select a subset of TRPs that provide reliable and informative measurements.

The selection of TRPs is performed based on the following criteria:
\begin{itemize}
    \item \textit{Geometric diversity:} TRPs with sufficiently separated azimuth angles relative to the target are preferred to avoid correlated LOS measurements.
    
    \item \textit{Detection reliability:} TRPs that consistently report detections (i.e., provide valid $\widehat{\mathbf{m}}_i$) across multiple CPIs are prioritized.
    
    \item \textit{Signal quality:} TRPs with higher SNR values $\gamma_i$ provide more reliable position and radial velocity estimates and are therefore preferred.
\end{itemize}

While incorporating a larger number of TRPs may improve robustness through spatial diversity, it also imposes a stricter multi-TRP voting requirement during the detection stage. This can increase the miss detection probability (MDP) when some TRPs fail to detect the target. This trade-off motivates the need for an appropriate voting strategy that balances detection reliability with robustness to missed observations.

\subsection{Target Association, Detection, and Parameter Estimation}

Given the set of measurements $\{\widehat{\mathbf{m}}_i\}$ from the selected TRPs, the next step is to determine which detections correspond to the same physical target and to estimate the target parameters in a consistent manner.

\subsubsection{Target Association via Spatial Gating}

Target association across TRPs is performed in Cartesian space using position-based clustering. All detections from different TRPs are aggregated and grouped using a spatial gating mechanism. Specifically, detections whose Euclidean distance is smaller than a predefined threshold $d_{3D}$ are considered to belong to the same candidate cluster. To prevent ambiguous associations, a unique assignment constraint is enforced such that each detection is assigned to at most one cluster. In cases where a detection satisfies the gating condition for multiple clusters, it is assigned to the cluster with the minimum Euclidean distance from the detection to the cluster centroid. This ensures that no detection contributes to multiple clusters.

\subsubsection{Detection via Multi-TRP Voting}

Each candidate cluster is validated using a multi-TRP voting mechanism. A cluster is declared as a valid target only if its constituent detections originate from at least $v_{\text{th}}$ distinct TRPs, where $v_{\text{th}}$ denotes the voting threshold. This voting process suppresses spurious detections arising from noise, clutter, and multipath-induced ghost targets, while improving overall detection reliability.

\subsubsection{Parameter Estimation via Fusion}

For each confirmed target, parameter estimation is performed using the associated multi-TRP measurements. The fused target position is obtained using power-weighted averaging of the individual TRP estimates: $\hat{\mathbf{p}} = \sum_{i} w_i \mathbf{p}_i, 
$
where  $\mathbf{p}_i = [x_i, y_i, z_i]^T$, $w_i = \frac{\gamma_i}{\sum_j \gamma_j}$, and $\gamma_i$ is the SNR of the $i$th detection. Each TRP provides a radial velocity measurement given by $v_{r,i} = \mathbf{v} \cdot \hat{\mathbf{u}}_i,$ where $\mathbf{v}$ is the true 3D velocity vector and $\hat{\mathbf{u}}_i$ is the unit LOS vector from TRP $i$ to the target. Stacking the measurements from multiple TRPs yields $\mathbf{v}_r = \mathbf{R}\mathbf{v},$
where $\mathbf{v}_r = [v_{r,1}, v_{r,2}, \ldots]^T$ and $\mathbf{R} \in \mathbb{R}^{N_{\mathrm{TRP}} \times 3}$ contains the stacked transposes of the LOS unit vectors. The velocity is estimated using a least-squares solution $\hat{\mathbf{v}} = \mathbf{R}^{\dagger}\mathbf{v}_r,$ where $(\cdot)^{\dagger}$ denotes the Moore–Penrose pseudo-inverse. Finally, the fused radial velocity with respect to a reference TRP (e.g., the center TRP in the deployment) is computed as $v_{r,\text{center}} = \hat{\mathbf{v}} \cdot \hat{\mathbf{u}}_{\text{center}},$ where $\hat{\mathbf{u}}_{\text{center}}$ is the LOS unit vector from the reference TRP to the estimated target position.

\subsection{Multi-TRP Sensing Resource and Overhead} 
\label{Sec:overhead}
In the multi-TRP framework, sensing resources are allocated to avoid mutual interference among the $P_{\mathrm{TRP}}$ participating TRPs. Assuming a 5G NR frame structure with $L_{\mathrm{sym}}$ OFDM symbols per slot (typically 14), let $L_{\mathrm{occ}}$ denote the total number of OFDM symbols occupied by sensing signals across all assisting TRPs within a single sensing occasion. The instantaneous resource overhead within a CPI, denoted as $\eta_{\mathrm{CPI}}$, is defined as the ratio of occupied sensing symbols to the total available symbols within the corresponding time interval, i.e., $\eta_{\mathrm{CPI}} = \frac{L_{\mathrm{occ}}}{L_{\mathrm{sym}}}$. To support scalable network deployment, the sensing results are updated every $T_{\mathrm{refresh}}$ seconds. Given a CPI duration of $T_{\mathrm{CPI}}$, the effective network sensing overhead, $\eta_{\mathrm{eff}}$, is obtained by scaling the instantaneous overhead by the sensing duty cycle, i.e., $\eta_{\mathrm{eff}} = \eta_{\mathrm{CPI}} \times \frac{T_{\mathrm{CPI}}}{T_{\mathrm{refresh}}}$.

\begin{figure*}[ht!]
\begin{minipage}[t]{0.25\textwidth}
    \centering
    \includegraphics[width=\linewidth]{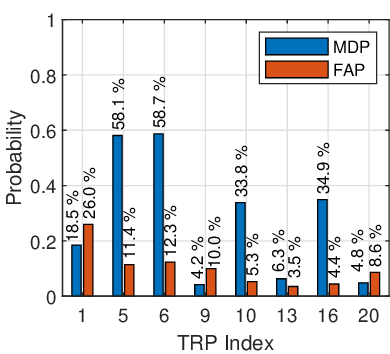}
    \panelcaption{(a) MDP and FAP}
\end{minipage}
\hspace{-0.25cm}
\begin{minipage}[t]{0.25\textwidth}
    \centering
    \includegraphics[width=\linewidth]{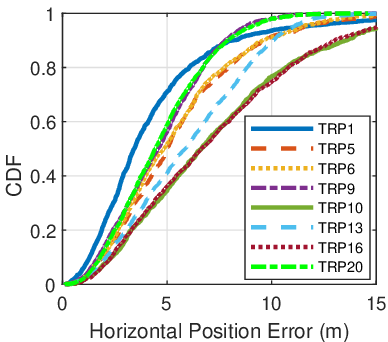}
    \panelcaption{(b) Horizontal position error}
\end{minipage}
\hspace{-0.3cm}
\begin{minipage}[t]{0.25\textwidth}
    \centering
    \includegraphics[width=\linewidth]{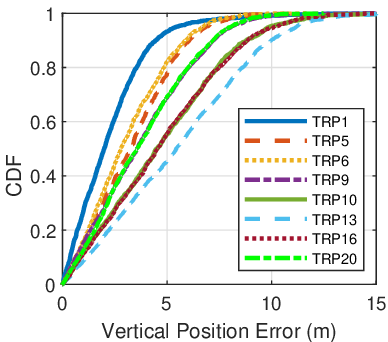}
    \panelcaption{(c) Vertical position error}
\end{minipage}
\hspace{-0.3cm}
\begin{minipage}[t]{0.25\textwidth}
    \centering
    \includegraphics[width=\linewidth]{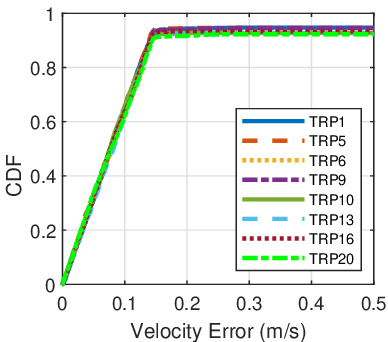}
    \panelcaption{(d) Radial velocity error}
\end{minipage}
\vspace{-0.2cm}
\caption{Detection, localization, and velocity parameter estimation performance of individual TRPs.}
\vspace{-0.5cm}
\label{fig:detection_TRPs}
\end{figure*}

\section{Evaluation Results and Insights}
\label{Sec:EvaluationResults}
In this section, we present system-level evaluation results obtained using our open-source NR ISAC simulation framework~\cite{nist_5gnrad}. The evaluation is conducted under the 3GPP UMa-AV scenario and follows the Release-19 ISAC channel modeling assumptions \cite{3GPP38901}. We consider a macro-cell TRP operating in a monostatic sensing configuration at a carrier frequency of $f_c = 4$~GHz. The system bandwidth is $B = 100$~MHz with subcarrier spacing $\Delta f = 30$~kHz (NR numerology $\mu=1$). Sensing utilizes downlink NR PRS signals transmitted on a CP-OFDM waveform. The PRS configuration uses a comb size of $K=2$ with $L_{\mathrm{PRS}}=2$ PRS symbols per occasion and a repetition period of $T_{\mathrm{PRS}}=1$~ms. Doppler estimation is performed over a CPI composed of $N_{\mathrm{CPI}}=128$ PRS occasions, resulting in a CPI duration of $T_{\mathrm{CPI}}=128$~ms.

The TRP is equipped with an $8\times8$ URA with $\pm45^\circ$ dual-polarized elements and a full-digital architecture. Two antenna spacings are considered, namely $(d_H,d_V)=(0.5,0.5)\lambda$ and $(0.5,0.8)\lambda$. The element radiation characteristics follow the 3GPP TR~38.901 antenna model \cite{3GPP38901}, providing a maximum element gain of $8$~dBi and a $65^\circ$ half-power beamwidth in both azimuth and elevation. The maximum transmit power at the TRP is set to $52$~dBm\footnote{dBm is unit of power level expressed using a logarithmic decibel (dB) scale respective to one milliwatt (mW).}, unless otherwise stated.

\begin{figure}[t]
\centering
\begin{minipage}[t]{0.24\textwidth}
    \centering
    \includegraphics[width=\linewidth]{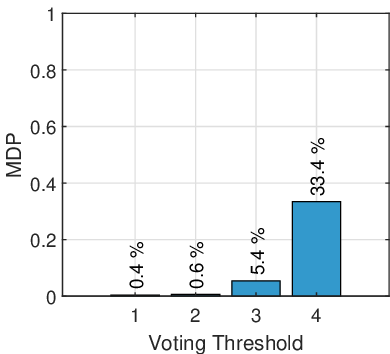}
    \panelcaption{(a) MDP versus $v_{\text{th}}$}
\end{minipage} 
\begin{minipage}[t]{0.24\textwidth}
    \centering
    \includegraphics[width=\linewidth]{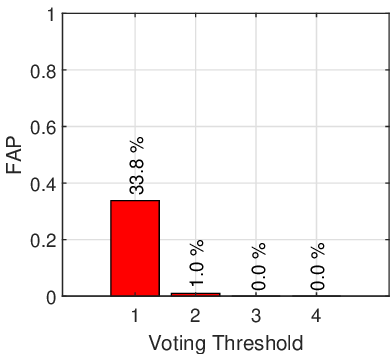}
    \panelcaption{(b) FAP versus $v_{\text{th}}$}
\end{minipage}
\vspace{-0.2cm}
\caption{Detection performance of the serving TRP1 with assistance from TRP9, TRP13, and TRP20.}
\label{fig:detection_TRP1_Coop}
\vspace{-0.5cm}
\end{figure}

\begin{figure*}[t]
\centering
\begin{minipage}[t]{0.25\textwidth}
    \centering
    \includegraphics[width=\linewidth]{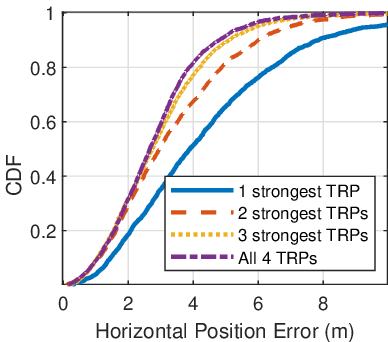}
    \panelcaption{(a) Horizontal position error}
\end{minipage}
%\hfill
\hspace{-0.25cm}
\begin{minipage}[t]{0.25\textwidth}
    \centering
    \includegraphics[width=\linewidth]{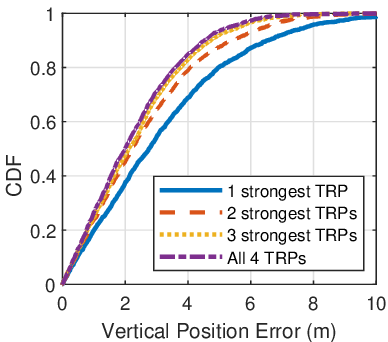}
    \panelcaption{(b) Vertical position error}
\end{minipage}
%\hfill
\hspace{-0.25cm}
\begin{minipage}[t]{0.25\textwidth}
    \centering
    \includegraphics[width=\linewidth]{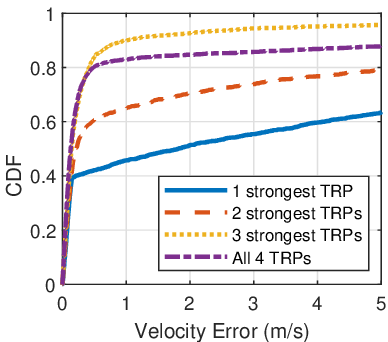}
    \panelcaption{(c) Radial velocity error}
\end{minipage}
\hspace{-0.25cm}
\begin{minipage}[t]{0.25\textwidth}
    \centering
    \includegraphics[width=\linewidth]{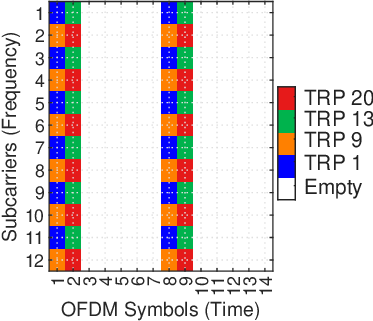}
    \panelcaption{(d) Sensing resource allocation}
\end{minipage}
\vspace{-0.2cm}
% \caption{ CDF comparison of localization and velocity estimation performance of serving TRP1 with multi-TRP assistance.}
\caption{ Performance and resource allocation for multi-TRP assisted sensing.}
\label{fig:CDF_TRP1_Coop}
\vspace{-0.2cm}
\end{figure*}
The evaluation results are obtained by averaging over $400$ independent simulation drops. For each simulation drop, $Q{=}5$ small UAV sensing targets are placed within the forward sector coverage region of TRP1 shown in Fig.~\ref{fig:deployment}. Their horizontal positions are uniformly sampled within the sector area, while their altitudes are uniformly distributed in the range from $25$~m to $300$~m. A minimum 3D distance of $10$~m is enforced between the TRP and each target, and the same minimum separation is maintained between targets. The horizontal velocity of each UAV is randomly selected within $[0,180]$~km/h with a random heading, while vertical motion is not considered. The RCS of the UAV targets follows the 3GPP ISAC model. The mean RCS is set to $\sigma_M{=}-12.81$~dBsm\footnote{dBsm is a unit of RCS expressed using a logarithmic decibel (dB) scale relative to one square meter}, and the small-scale fluctuation component is modeled as a log-normal random variable with standard deviation $\sigma_{\sigma_S}{=}3.74$~dB. The target propagation channel is generated according to the Release-19 sensing target model \cite{3GPP38901}, where multipath components with power levels more than $40$~dB below the strongest path are removed, consistent with the 3GPP evaluation methodology \cite{3GPP38765}. In addition to target echoes, a monostatic background channel is included to represent clutter multipath in the environment, generated according to the Release-19 background channel procedure. The detection reliability is quantified using two primary metrics:  MDP and false alarm probability (FAP) \cite{3GPP38765}. The MDP is calculated as the average conditional probability that a target remains undetected despite being present in the simulation environment, defined as:  MDP = $\sum_{n=0}^{N-1} \frac{D_n}{M_n}/N$ where $D_n$ denotes the count of targets in drop $n$ that failed to associate with any detected object, $M_n$ is the total number of ground-truth targets in that drop, and $N$ represents the aggregate number of drops containing at least one target. Complementarily, the FAP accounts for instances where a detected object cannot be mapped to any ground-truth target. In scenarios where true targets are present, the FAP is expressed as: FAP = $\frac{1}{K} \sum_{\substack{0 \leq n < N, \\ M_n' \neq 0}} \frac{D_n'}{M_n'}$. Here, $D_n'$ is the number of "ghost" or unassociated detections in drop $n$, $M_n'$ is the total number of reported objects, and $K$ is the subset of drops where at least one detection occurred. This approach ensures that the performance metrics reflect the system's ability to distinguish between legitimate aerial targets and environmental clutter within the 3GPP-specified UMa-AV region.

% \begin{figure*}[t]
% \centering
% \begin{minipage}[t]{0.3\textwidth}
%     \centering
%     \includegraphics[width=\linewidth]{figures/TRP1-HorizontalPositionError.png}
%     \caption*{(a) Horizontal position error}
% \end{minipage}
% \hfill
% \begin{minipage}[t]{0.3\textwidth}
%     \centering
%     \includegraphics[width=\linewidth]{figures/TRP1-VerticlePositionError.png}
%     \caption*{(b) Vertical position error}
% \end{minipage}
% \hfill
% \begin{minipage}[t]{0.3\textwidth}
%     \centering
%     \includegraphics[width=\linewidth]{figures/TRP1-VelocityError.png}
%     \caption*{(c) Radial velocity error}
% \end{minipage}
% \caption{CDF comparison of localization and velocity estimation performance of serving TRP 1 with no cooperation.}
% \label{fig:CDF_TRP1_NoCoop}
% \end{figure*}

To study multi-TRP assisted sensing performance, we initially identify a candidate set of eight TRPs (1, 5, 6, 9, 10, 13, 16, and 20) situated in the immediate vicinity of the target region. As illustrated in the network layout in Fig.~\ref{fig:deployment}, these TRPs provide the primary spatial coverage for the designated UAV sensing area. Fig.~\ref{fig:detection_TRPs}(a) shows the MDP and FAP obtained for these candidate TRPs considering $(d_H,d_V)=(0.5,0.5)\lambda$. The results reveal significant variability across TRPs due to differences in geometry, sector FOV, and propagation conditions. For instance, TRPs 5 and 6 exhibit high miss detection probabilities (around 58~\%), while TRPs 10 and 16 also show relatively high MDP values above 30~\%. In contrast, TRPs 9, 13, and 20 demonstrate consistently low miss detection probabilities (below 7~\%) with moderate false alarm levels. Although TRP1 shows slightly higher MDP and FAP compared to some neighboring nodes, it provides accurate target position and radial velocity estimates whenever the target is successfully detected as shown in Figure \ref{fig:detection_TRPs}(b-d). 

Including all eight TRPs would improve sensing robustness but would also increase sensing signaling and processing overhead. Therefore, a down-selection strategy is applied based on the three criteria described in the TRP selection procedure, namely detection reliability, geometric diversity, and signal quality. Based on these criteria along with performance as shown in Fig.~\ref{fig:detection_TRPs}, four TRPs (1, 9, 13, and 20) are selected as assisting nodes due to their favorable spatial distribution around the target and consistent detection performance in the considered scenario. Although TRP1 and TRP13 have similar sector orientations and azimuth FOV, TRP13 is included to provide redundancy when TRP1 fails to detect the target due to limited vertical coverage at high-elevation angles. Other nearby TRPs either fail to reliably detect the UAV or provide measurements with limited geometric diversity, making them less suitable for sensing. By retaining only these four TRPs, the system achieves the performance gains required to meet 3GPP objectives while minimizing the network resource consumption quantified in Section~\ref{Sec:overhead}. In a practical deployment, the set of four assisting TRPs can be dynamically adapted using real time sensing quality metrics and prevailing network conditions, with candidate TRPs retained only if their SNR $\gamma_i$ exceeds a predefined threshold.     

Fig.~\ref{fig:detection_TRP1_Coop} shows the detection performance of the serving TRP1 with multi-TRP-assisted sensing using TRP9, TRP13, and TRP20 under different voting thresholds, with $d_{3D}=20$~m. As shown in the figure, increasing the voting threshold significantly reduces the FAP, but this improvement comes at the expense of a higher MDP, highlighting the inherent detection trade-off in multi-TRP voting schemes. When the voting threshold is set to one, the system achieves very low MDP (0.4~\%) but suffers from a high FAP (33.8~\%). Conversely, increasing the threshold to four effectively eliminates false alarms, but results in a substantial increase in missed detections (33.4~\%). A voting threshold of two provides a balanced operating point, achieving near-zero FAP while maintaining a low MDP of approximately 0.6~\%. Therefore, a voting threshold of two is used in the subsequent multi-TRP assisted sensing evaluations.

Once a target is confirmed using a voting threshold of two, the network can further refine the target position and velocity estimation by selecting the $K$ strongest TRPs--defined as those reporting the highest estimated SNR for the associated detection--from the four assisting nodes. As illustrated by the {cumulative distribution function (CDF)} results in Fig.~\ref{fig:CDF_TRP1_Coop}, the horizontal and vertical position errors consistently decrease as the number of assisting TRPs increases from one to four. This improvement is expected since incorporating measurements from additional TRPs reduces the spatial uncertainty region and enhances geometric diversity for position estimation. In contrast, radial velocity estimation exhibits a different behavior. While utilizing four TRPs with a voting threshold of two ensures robust target confirmation, the velocity error is further optimized when using only the three strongest TRPs. Including a fourth TRP results in a degradation in velocity accuracy, which can be attributed to a measurement with lower SNR or unfavorable geometry that increases the sensitivity to noise during the least-squares reconstruction of the 3D velocity.
\begin{table*}[h]
\centering
\caption{Performance for serving TRP with/ without assistance at 4 GHz under different transmit powers and antenna spacings}
\label{tab:serving_trp_4ghz_with_cooper}
\renewcommand{\arraystretch}{1.15}
\setlength{\tabcolsep}{4pt}
\begin{tabular}{|l|c|c|c|c|c|c|c|c|c|}
\hline
\multirow{3}{*}{\textbf{Performance Metric}} & \multirow{3}{*}{\shortstack{\textbf{3GPP} \\ \textbf{Requirement}}} & \multicolumn{4}{c|}{\textbf{52~dBm}} & \multicolumn{4}{c|}{\textbf{37~dBm}} \\ \cline{3-10} 
 &  & \multicolumn{2}{c|}{No Assistance} & \multicolumn{2}{c|}{With Assistance} & \multicolumn{2}{c|}{No Assistance} & \multicolumn{2}{c|}{With Assistance} \\ \cline{3-10} 
 &  & $(0.5, 0.5)\lambda$ & $(0.5, 0.8)\lambda$ & $(0.5, 0.5)\lambda$ & $(0.5, 0.8)\lambda$ & $(0.5, 0.5)\lambda$ & $(0.5, 0.8)\lambda$ & $(0.5, 0.5)\lambda$ & $(0.5, 0.8)\lambda$ \\ \hline
Horizontal error @90~\% [m] & 10 & 8.5 & 10 & 6 & 6.3 &10.75 & 11.75 &6 & 6.32\\ 
Vertical error @90~\% [m] & 10 & 4.25 & 3.2 & 5 & 3.5 & 6.15& 4.35&5.5 & 4\\ 
Velocity error @90~\% [m/s] & 5 & 0.15 & 0.15 & 0.95 & 2 & 0.15 & 6.75&2.12 & 4.75\\ 
Miss detection [\%] & 5 & 18.5 & 57.8 & 0.6 & 1.4 &18.8 & 57.2&1.5 & 2.6 \\ 
False alarm [\%] & 5 & 26 & 35.5 & 1 & 1 & 31&41.4 & 2.1& 1.3 \\ \hline
\end{tabular}
\vspace{-0.5cm}
\end{table*}

Table~\ref{tab:serving_trp_4ghz_with_cooper} summarizes the quantitative evaluation metrics extracted from the experiments for $(0.5, 0.5)\lambda$ and $(0.5, 0.8)\lambda$ antenna spacings under different transmit powers, alongside the corresponding 3GPP requirements for each performance indicator. Overall, multi-TRP assisted sensing -- utilizing a voting threshold of two for target confirmation and least-squares velocity reconstruction with the three strong TRPs -- substantially enhances detection reliability while maintaining high-accuracy position and velocity estimates that consistently meet or exceed 3GPP requirements. For example, at a transmit power of 52~dBm with $(0.5, 0.5)\lambda$ spacing, switching from a single-TRP configuration to assistance improves the 90~\% horizontal localization error from 8.5~m to 6~m. Most notably, assistance with a voting threshold of two  reduces the MDP from 18.5~\% to 0.6~\% and suppresses the FAP from 26~\% to 1~\%. The benefits of multi-TRP assisted sensing are even more pronounced at a lower transmit power of 37~dBm.  While assistance introduces a slight increase in the 90~\% velocity error compared to the highly optimistic single-TRP detections, the resulting values in the range (0.95 to 2)~m/s remain well within the 5~m/s 3GPP limit, demonstrating that multi-TRP fusion provides a robust and standardized-compliant solution for UAV sensing. Comparing the two antenna configurations, increasing the vertical element spacing from $0.5\lambda$ to $0.8\lambda$ leads to poorer detection performance--evidenced by the sharp increase in miss detection rates--yet multi-TRP assistance remains resilient enough to meet 3GPP requirements even at the lower $37$~dBm transmit power.

Finally, we compute the sensing signal overhead according to 3GPP agreement\cite{3GPP38765}. As illustrated in the resource allocation grid in Fig.~\ref{fig:CDF_TRP1_Coop}(d), when $P_{\mathrm{TRP}}=4$ TRPs perform sensing collaboratively using frequency-division multiplexing, they occupy $L_{\mathrm{occ}}=4$ symbols out of the $L_{\mathrm{sym}}=14$ available symbols in the slot. This results in an instantaneous overhead of $\eta_{\mathrm{CPI}} \approx 29~\%$ during the active sensing period.  However, this overhead is significantly mitigated by the temporal scheduling of sensing occasions. By utilizing a coherent processing interval of $T_{\mathrm{CPI}} = 128$~ms and extending the refresh interval to $T_{\mathrm{refresh}} = 1$ s, the effective network overhead $\eta_{\mathrm{eff}}$ is reduced to approximately 3.7~\%. This result demonstrates that multi-TRP assisted UAV sensing can be implemented in a scalable and resource-efficient manner that remains compatible with concurrent communication services.

% Finally, we also compute the sensing signal overhead based on the 3GPP agreement \cite{3GPP38765}. The evaluation of sensing overhead for multi-TRP assisted sensing (specifically TRPs 1, 9, 13, and 20) relies on the 3GPP NR ISAC resource definitions, focusing exclusively on Type 1 resources for signal transmission and Type 3 resources that are muted for communications to prevent interference during sensing operations. For each TRP, the resource allocation utilizes a Comb-2 pattern in the frequency domain with two OFDM symbols per occasion. As illustrated in the Multi-TRP resource allocation in Fig.~\ref{fig:PRS}, 
% when four TRPs perform sensing collaboratively, they utilize frequency-division multiplexing to share sensing symbols; for example, TRP1 and TRP9 are multiplexed on OFDM symbols 1 and 8, while TRP13 and TRP20 occupy symbols 2 and 9. Under this configuration, the sensing overhead during a single CPI is calculated at 29\%. However, this overhead can be significantly reduced by extending the refresh interval beyond the 128 ms CPI duration; by updating sensing results every 1 s, the effective network overhead is lowered to approximately 3.7\%, supporting the scalable and resource-efficient deployment of multi-TRP assisted UAV sensing.
% \begin{figure}[t!]
% \centering
% \includegraphics[width=0.6\linewidth]{figures/PRS.png}
% \caption{Sensing resource allocation for multi-TRP assisted sensing.}
% \label{fig:PRS}
% \vspace{-0.5cm}
% \end{figure}
\section{Conclusion}
\label{Sec:Conclusion}
This paper examined the impact of multi-TRP assistance for UAV detection in a 3GPP NR ISAC framework. While single TRP sensing suffers from high miss detection rates due to limited motion observability, multi-TRP assisted sensing significantly improves detection performance and localization robustness, while enabling the reconstruction of the targets’ true 3D velocity. By implementing a clustering-based association combined with a voting mechanism and least-squares velocity reconstruction, the system ensures consistent fusion performance. Our evaluation reveals that positioning accuracy consistently improves when adding up to four TRPs, whereas radial velocity estimation is optimized by utilizing the three strongest TRPs to avoid noise from less favorable geometries. Among the evaluated configurations, a voting threshold of two TRPs was identified as providing the best trade-off between detection reliability and false alarm suppression. Furthermore, the study demonstrates that sensing overhead can be reduced to approximately 3.7\% when refresh intervals are extended beyond the CPI duration, supporting the practical and resource-efficient deployment of ISAC in urban macro environments.  These findings establish multi-TRP assistance as a necessary architectural component for reliable NR-based UAV sensing in urban environments and provide concrete design guidelines for future 3GPP ISAC standardization. Future studies should investigate asynchronous fusion in the presence of backhaul delay by explicitly accounting for UAV motion over finite time intervals.

\bibliographystyle{IEEEtran}
\bibliography{reference}

\end{document}